\documentclass[twocolumn,         
               showpacs,longbibliography,            
               showkeys,preprintnumbers,     
               aps,                 
               prd,          	    
               letterpaper,             
               superscriptaddress,      
               nofootinbib,         
               tightenlines,        
               floats,floatfix      
               ]{revtex4-1}
\usepackage{physics}
\usepackage{epsf}
\usepackage{graphicx,color}
\usepackage{subfigure}
\usepackage{latexsym}
\usepackage{amsmath,amssymb}        
\usepackage[colorlinks=true,linkcolor=blue,citecolor=blue]{hyperref}
\usepackage{mathrsfs}
\usepackage{comment}
\usepackage{soul}
\usepackage{cancel}
\definecolor{purple}{rgb}{0.58,0.0,0.83}
\definecolor{orange}{rgb}{1,0.5,0}
\DeclareSymbolFontAlphabet{\mathrsfs}{rsfs}
\DeclareMathAlphabet{\mathcal}{OMS}{cmsy}{m}{n}

\begin{document}


\title{Kinematic Imprints of vortex-lines of BEC Dark Matter on Baryonic Matter}


\author{Iv\'an  \'Alvarez-Rios}
\email{ivan.alvarez@umich.mx}
\affiliation{Instituto de F\'{\i}sica y Matem\'{a}ticas, Universidad
              Michoacana de San Nicol\'as de Hidalgo. Edificio C-3, Cd.
              Universitaria, 58040 Morelia, Michoac\'{a}n,
              M\'{e}xico.}     

\author{Carlos Tena-Contreras}
\email{1423327c@umich.mx}
\affiliation{Instituto de F\'{\i}sica y Matem\'{a}ticas, Universidad
              Michoacana de San Nicol\'as de Hidalgo. Edificio C-3, Cd.
              Universitaria, 58040 Morelia, Michoac\'{a}n,
              M\'{e}xico.}           

\author{Francisco S. Guzm\'an}
\email{francisco.s.guzman@umich.mx}
\affiliation{Instituto de F\'{\i}sica y Matem\'{a}ticas, Universidad
              Michoacana de San Nicol\'as de Hidalgo. Edificio C-3, Cd.
              Universitaria, 58040 Morelia, Michoac\'{a}n,
              M\'{e}xico.}


\date{\today}


\begin{abstract}
Our results demonstrate that vortex lines in Bose-Einstein Condensate Dark Matter (BECDM) can act as gravitational seeds that induce the condensation of baryonic matter, leading to localized gas accumulation even in the absence of imposed symmetries or rotation. 
Our analysis is based on the numerical solution of the system of equations for the BECDM gravitationally coupled to Euler equations for a compressible ideal gas (IG) that we use as a model of baryonic matter. Numerical simulations are constructed for various scenarios that start with a vortex solution for the BECDM and a randomly distributed ideal gas, with the aim of investigating whether the matter distribution and dynamics of the vortex influences the dynamics and distribution of the gas.
We find that the gas condensation process is most efficient when the IG mass dominates over the BECDM, and when the IG has low initial velocity dispersion. 
We also find that strong bosonic self-interaction does not guarantee the vortex stability, instead, it can trigger dynamical instabilities that disrupt both the vortex structure and the surrounding gas. 
An interesting finding is that the vortex drives a persistent morphological signature on the gas, often in the form of ring-like features visible in projected density maps. These patterns survive nonlinear evolution and may serve as indirect tracers of vortex structures in BECDM halos, potentially offering a novel and testable observational probe of the model.
\end{abstract}




\maketitle

\section{Introduction}
\label{sec:intro}

Bose–Einstein condensates (BECs) are quantum-coherent states of matter that arise at ultra-low temperatures, where a large number of bosons occupy the ground state. A hallmark of rotating BECs is the formation of quantized vortices, which are topological defects characterized by vanishing density at their cores and quantized circulation of the wave function. These vortices have been extensively observed in ultracold atomic gas experiments, where they exhibit a variety of dynamical behaviors, including nucleation, lattice formation, and complex interactions governed by quantum hydrodynamics~\cite{PethickSmith, Fetter:2009, Matthews1999, Anderson2001, Leanhardt2002, Sadler_2006, Schweikhard2004}. Early experiments demonstrated vortex creation via phase imprinting~\cite{Leanhardt2002}, angular momentum transfer~\cite{Matthews1999}, and spontaneous generation through dynamical instabilities such as soliton decay~\cite{Anderson2001}. These phenomena have significantly advanced our understanding of superfluidity and topological excitations in quantum fluids.

In recent years, the idea that dark matter may consist of a self-gravitating Bose-Einstein condensate has attracted increasing attention. In the Bose-Einstein Condensate Dark Matter (BECDM), dark matter is composed of ultralight bosons with masses around \( m_B \sim 10^{-22} \, \mathrm{eV} \), whose dynamics are governed by the Gross-Pitaevskii-Poisson (GPP) system~\cite{Matos-Urena:2000, Sahni:2000, Hu:2000, Hui:2016}. This model addresses various of the potential small-scale challenges faced by the Cold Dark Matter (CDM) paradigm, including the cusp-core discrepancy in galactic centers~\cite{Marsh-Pop:2015, Harko:2011xw}, the overabundance of satellite galaxies~\cite{Klypin:1999, Schive:2014dra}, and the suppression of small-scale structures due to quantum pressure~\cite{Hui:2016}.

Due to its wave-like nature, BECDM exhibits phenomena analogous to those in condensed matter systems, including the formation of quantized vortices under rotation. These astrophysical vortices macroscopic counterparts of those found in laboratory BECs are topological defects with vanishing density and quantized phase winding. Their formation and dynamics have been explored in simulations of rotating FDM halos, where they are found to influence core structure, angular momentum transport, and global halo morphology~\cite{Mocz:2017wlg, Nikolaieva_2021, Nikolaieva_2023, Alexander:2021zhx}.

The conditions under which such structures can form have been studied in~\cite{Rindler_Daller_2012, Schobesberger:2021ghi}, showing that vortex formation in BECDM halos requires a repulsive self-interaction. In particular, configurations with vanishing self-interaction do not support vortex formation under realistic astrophysical conditions in galactic nuclei. The long-term stability of vortex lines has been confirmed in~\cite{Nikolaieva_2021, Nikolaieva_2023}, demonstrating that these structures can persist over cosmological timescales. Similar conclusions have been obtained within the Thomas--Fermi approximation~\cite{Kain_2010}, and further studies have extended the stability analysis to vortices in the vicinity of black holes~\cite{Glennon:2023oqa}.

Beyond their formation and long-term stability, recent studies have begun to explore the nonlinear dynamics of vortices and solitons in self-gravitating BECDM systems. In particular, numerical investigations of head-on and off-center mergers between condensates~\cite{sivakumar2025revealingturbulentdarkmatter} have revealed a rich phenomenology, including soliton-mediated instabilities that give rise to turbulence with scale-dependent energy cascades. This turbulence is initially sustained by the decay of dark solitons into vortex structures. Unlike turbulence in harmonically trapped atomic condensates, the self-gravitating nature of BECDM significantly modifies the compressible flow spectrum and the morphology of density waves, underscoring the critical role of gravity in structuring the condensate. These findings highlight the importance of vortex and soliton dynamics as potential probes of BECDM phenomenology, particularly in astrophysical scenarios involving mergers, relaxation processes, or halo reconfigurations.

The potential astrophysical relevance of vortices has motivated efforts to identify observational signatures that could constrain the properties of the underlying bosonic field. Vortex lines may leave measurable imprints on galactic rotation curves, central density profiles, and the dynamical evolution of baryons, offering a novel window into the mass of the dark matter particle and the angular momentum content of halos. Features such as central density depressions or interference-induced granularity, observable in high-resolution data from dwarf and low surface brightness galaxies, have been proposed as potential indirect evidence for the quantum nature of dark matter~\cite{RindlerDaller:2013, Schive:2014hza, Mocz:2017wlg}.

Beyond individual halo dynamics, the collective behavior of vortex structures in BECDM may have important implications for structure formation and galaxy evolution. In analogy with turbulence in quantum fluids, a cosmological distribution of vortices could seed anisotropies, drive internal heating of baryonic components, or even influence feedback mechanisms during star formation. Moreover, interactions between vortices and large-scale gravitational potentials could result in angular momentum redistribution or fragmentation processes not accounted for in classical dark matter models. As such, vortex dynamics in BECDM represent a fertile ground for connecting fundamental properties with observable phenomena.

From an observational perspective, such dynamical signatures could manifest as localized perturbations in rotation curves, non-circular motions, or anisotropic gas flows particularly in dwarf and low surface brightness (LSB) galaxies, where the baryonic component is dynamically subdominant and more susceptible to subtle gravitational effects. High-resolution kinematic surveys, such as those from the THINGS~\cite{Walter_2008} and LITTLE THINGS~\cite{Hunter_2012} collaborations, offer promising avenues to detect these features. Identifying such imprints would provide a novel observational probe into the internal dynamics of BECDM-dominated halos and help constrain the quantum properties of dark matter.

Moreover, certain oscillation modes of BECDM structures have been shown to transfer gravitational energy to the baryonic component~\cite{Guzman2019, AlvarezGuzman2022, Alvarez_Rios_2023, alvarezrios2024fermionbosonstarsattractorsfuzzy}. This suggests that vortex-bearing BECDM configurations, through their toroidal gravitational potentials, could induce distinctive dynamical responses in the baryonic gas. In this framework, baryonic matter modeled as an ideal gas (IG) governed by the Euler equations may serve as an indirect tracer of underlying quantum-coherent structures such as vortices.

Inspired by these possibilities, in this paper we solve the evolution equations of the BECDM coupled gravitationally to an ideal gas, which as a first approximation can be used to model luminous matter and thus can act as a tracer of the potentially existing BECDM vortex-related structures. We solve this coupled system in order to investigate the stability of vortices, the influence of the self-interacting coupling of the bosons, the effects on stability of BECDM or IG dominated scenarios and the structures formed by the IG due to the vortex dynamics.

This article is organized as follows. In Section~\ref{sec:model}, we present the governing equations along with the mathematical and numerical framework used to carry out the simulations. Section~\ref{sec:results} discusses the results and their physical implications. Finally, Section~\ref{sec:conclusions} summarizes our conclusions.

\section{Modeling a Self-Consistent Vortex in BECDM}
\label{sec:model}

\subsection{Model and Equations}

The dynamics of BECDM, gravitationally coupled to an IG, is described by the Gross-Pitaevskii-Poisson-Euler (GPPE) system~\cite{AlvarezGuzman2022, Alvarez_Rios_2023, alvarezrios2024fermionbosonstarsattractorsfuzzy}:

\begin{eqnarray}
i\hbar \, \partial_{\tilde{t}} \tilde{\Psi} = - \frac{\hbar^2}{2 m_B} \nabla_{\tilde{x}}^2 \tilde{\Psi} + m_B \tilde{V} \tilde{\Psi} + \tilde{g} |\tilde{\Psi}|^2 \tilde{\Psi}, \label{eq:Schrodinger} \\
\partial_{\tilde{t}} \tilde{\rho} + \nabla_{\tilde{x}} \cdot (\tilde{\rho} \, \vec{\tilde{v}}) = 0, \label{eq:mass_conservation} \\
\partial_{\tilde{t}} (\tilde{\rho} \, \vec{\tilde{v}}) + \nabla_{\tilde{x}} \cdot (\tilde{\rho} \, \vec{\tilde{v}} \otimes \vec{\tilde{v}} + \tilde{p} \, \mathbf{I}) = -\tilde{\rho} \nabla_{\tilde{x}} \tilde{V}, \label{eq:momentum_conservation} \\
\partial_{\tilde{t}} \tilde{E} + \nabla_{\tilde{x}} \cdot \left[ \vec{\tilde{v}} (\tilde{E} + \tilde{p}) \right] = -\tilde{\rho} \vec{\tilde{v}} \cdot \nabla_{\tilde{x}} \tilde{V}, \label{eq:energy_conservation} \\
\nabla_{\tilde{x}}^2 \tilde{V} = 4\pi G \left(\tilde{\rho}_T - \bar{\tilde{\rho}}_T\right), \label{eq:Poisson} \\
\tilde{p} = (\gamma - 1) \tilde{\rho} \tilde{e}. \label{eq:EoS}
\end{eqnarray}

\noindent
Here, $\tilde{\Psi}$ is the macroscopic wave function of the BECDM component, with mass density $m_B |\tilde{\Psi}|^2$. The parameter $m_B$ is the boson mass, $\hbar$ is the reduced Planck constant, and $\tilde{g} = 4\pi \hbar^2 \tilde{a}_s / m_B$ is the nonlinear self-interaction coefficient, with $\tilde{a}_s$ denoting the $s$-wave scattering length.

For the IG equations, $\tilde{\rho}$ is the mass density of each volume element, $\vec{\tilde{v}}$ its velocity field, 
$\tilde{E} = \tilde{\rho} \left( \tilde{e} + \frac{1}{2} |\vec{\tilde{v}}|^2 \right)$ the total energy with specific internal energy $\tilde{e}$ and  $\mathbf{I}$ represents the identity tensor. The pressure $\tilde{p}$ is related to $\tilde{e}$ through the ideal gas equation of state~\eqref{eq:EoS}, with adiabatic index $\gamma$.

The gravitational potential $\tilde{V}$ couples the condensate and gas components through the total density $\tilde{\rho}_T = \tilde{\rho} + m_B |\tilde{\Psi}|^2$ and its spatial average $\bar{\tilde{\rho}}_T$. The system is expressed in Cartesian coordinates $(\tilde{t}, \vec{\tilde{x}})$.

\subsection{Non-dimensionalization}

To eliminate physical units and facilitate numerical implementation, we adopt a dimensionless formulation, following the strategy in~\cite{AlvarezGuzman2022, CarlosIvanFranciscoUniverse, palomareschavez2024blackholescondensationpoints}. This is achieved by rescaling all variables as follows:
\[
\begin{aligned}
\tilde{t} &= t_0 \, t,            & \quad \vec{\tilde{x}} &= x_0 \, \vec{x},     & \quad \tilde{\Psi} &= \Psi_0 \, \Psi, \\
\tilde{\rho} &= \rho_0 \, \rho,   & \quad \vec{\tilde{v}} &= v_0 \, \vec{v},     & \quad \tilde{p} &= p_0 \, p, \\
\tilde{e} &= e_0 \, e,            & \quad \tilde{E} &= E_0 \, E,                 & \quad \tilde{V} &= V_0 \, V, \\
\tilde{g} &= g_0 \, g,            & \quad \tilde{a}_s &= a_0 \, a_s.             & \quad              &
\end{aligned}
\]

\noindent
Tilded quantities denote dimensional physical variables, while untilded ones are their dimensionless counterparts.

A natural choice of scaling parameters is based on the boson mass \( m_B = m_{22} \times 10^{-22} \, \text{eV}/c^2 \) and a characteristic length scale \( x_0 = \mathrm{kpc}/\lambda \). From these, the following scale factors are derived:

\begin{eqnarray}
t_0 &=& \dfrac{m_B x_0^2}{\hbar} \approx 50.96 \times 10^{-3} \left( \dfrac{m_{22}}{\lambda^2} \right) \, \mathrm{Gyr}, \\
v_0 &=& \dfrac{\hbar}{m_B x_0} \approx 19.20 \left( \dfrac{m_{22}}{\lambda} \right) \, \mathrm{km/s}, \\
M_0 &=& \dfrac{\hbar^2}{4 \pi G m_B^2 x_0} \approx 6.820 \times 10^6 \left( \dfrac{\lambda}{m_{22}^2} \right) \, M_\odot, \\
\rho_0 &=& \dfrac{\hbar^2}{4 \pi G m_B^2 x_0^4} \approx 6.820 \times 10^6 \left( \dfrac{\lambda^4}{m_{22}^2} \right) \, \dfrac{M_\odot}{\mathrm{kpc}^3}, \\
a_0 &=& \dfrac{G m_B^3 x_0^2}{\hbar^2} \approx 3.228 \times 10^{-75} \left( \dfrac{m_{22}^3}{\lambda^2} \right) \, \mathrm{cm}.
\end{eqnarray}

The remaining scaling factors, for pressure, energy, potential, and internal energy—can be derived consistently from these base units. Converting solutions back to physical units requires specifying both the dimensionless scale parameter \(\lambda\) and the boson mass parameter \(m_{22}\). The dimensionless GPPE system then becomes:

\begin{eqnarray}
i \, \partial_t \Psi = -\dfrac{1}{2} \nabla^2 \Psi + V \Psi + g |\Psi|^2 \Psi, \label{eq:Schrodinger_adim} \\
\partial_t \rho + \nabla \cdot (\rho \vec{v}) = 0, \label{eq:mass_adim} \\
\partial_t (\rho \vec{v}) + \nabla \cdot (\rho \vec{v} \otimes \vec{v} + p \mathbf{I}) = -\rho \nabla V, \label{eq:momentum_adim} \\
\partial_t E + \nabla \cdot \left[ \vec{v} (E + p) \right] = -\rho \vec{v} \cdot \nabla V, \label{eq:energy_adim} \\
\nabla^2 V = \rho_T - \bar{\rho}_T, \label{eq:Poisson_adim} \\
p = (\gamma - 1) \rho e. \label{eq:EoS_adim}
\end{eqnarray}

\noindent
This system defines the complete dimensionless GPPE framework used in our numerical simulations. We now proceed to describe the initial conditions.
\subsection{Initial Conditions}

The goal is to initialize a vortex line in the BECDM component and investigate its dynamical influence on the surrounding baryonic medium. To this end, we prescribe an initial wave function of the form suggested in~\cite{Nikolaieva_2021, Nikolaieva_2023}:
\begin{equation} 
    \Psi(0, \vec{x}) = \phi(r_{\perp}, z) \, e^{i m \varphi},
\end{equation}

\noindent
where \(m\) is the topological charge (or winding number) of the vortex, and \(\phi(r_{\perp}, z)\) is a real-valued amplitude profile that satisfies the stationary GPP equations under the boundary conditions \(\phi(0,z) = 0\), \(\partial_z \phi(r_{\perp},0) = 0\), and the vanishing of \(\phi\) and its derivatives at infinity (see Appendix~\ref{app:ground_state}). This profile is constructed in cylindrical coordinates \((r_{\perp}, \varphi, z)\).

This setup defines a self-consistent, rotating solitonic structure with a vortex line aligned along the \(z\)-axis, serving as a convenient initial condition for studying the interplay between angular momentum and the IG response in the coupled BECDM-baryon system. This wave function written in cylindrical coordinates, is later on interpolated into the three-dimensional space described with Cartesian coordinates, where the evolution will take place.

The initial conditions for the IG component are generated from an auxiliary wave function:
\begin{equation}
    \Psi_{\mathrm{aux}} = A \, \mathcal{F}^{-1} \left\lbrace e^{-|\vec{p}|^2} \, e^{i \Phi(\vec{p})} \right\rbrace,
\end{equation}

\noindent
where \(A\) is a normalization constant chosen such that the total gas mass equals \(M_{\mathrm{IG}}\). Here, \(\vec{p}\) denotes Cartesian momentum-space coordinates, and \(\Phi(\vec{p})\) is a random phase uniformly distributed in \([0, 2\pi)\) for each mode, following the prescription in~\cite{Rusos2018, Chen2021, alvarezrios2024fermionbosonstarsattractorsfuzzy}.

The initial gas density is given by \(\rho(0, \vec{x}) = |\Psi_{\mathrm{aux}}|^2\). The velocity field \(\vec{v}(0, \vec{x})\) is assigned randomly, subject to the constraint \( |\vec{v}(0, \vec{x})| \leq v_{\mathrm{m}} \), where \(v_{\mathrm{m}}\) represents the maximum allowed gas speed at initialization. The initial pressure is set using a polytropic equation of state:

\[
p(0, \vec{x}) = K \, \rho(0, \vec{x})^\gamma,
\]
where \(K\) is the polytropic constant, chosen to match the desired energy scale of the gas. The corresponding internal energy is computed from the ideal gas law:
\[
e(0, \vec{x}) = \frac{p(0, \vec{x})}{(\gamma - 1)\, \rho(0, \vec{x})} = \frac{K}{\gamma - 1} \, \rho(0, \vec{x})^{\gamma - 1},
\]
as defined in Eq.~\eqref{eq:EoS}.

\subsection{Numerical Setup}

The simulations are performed using the \textsc{CAFE-FDM} code, described in Refs.~\cite{AlvarezGuzman2022, periodicas}. The system of equations is solved in a 3D periodic domain described with Cartesian coordinates. All the evolution equations are integrated using a third-order Runge-Kutta scheme.

The Gross-Pitaevskii equation \eqref{eq:Schrodinger} is discretized using the Fast Fourier Transform (FFT), allowing for efficient and spectrally accurate computation of spatial derivatives.  Euler equations [\eqref{eq:mass_conservation}-\eqref{eq:energy_conservation}] are solved using high-resolution shock-capturing (HRSC) methods. In particular, we employ the Minmod linear reconstructor and the Harten-Lax-van Leer-Einfeldt (HLLE) approximate Riemann solver to compute numerical fluxes at cell interfaces.

Gravitational coupling between the BECDM and IG components is consistently enforced by solving the Poisson equation \eqref{eq:Poisson} at each RK3 substep using the FFT.

The system is evolved for 500 dimensionless time units, on the numerical domain, a cubic box of size \(L = 80\), discretized with \(N = 128\) grid points per dimension, yielding a uniform spatial resolution \(h = L/N\). The time step \(\Delta t\) is fixed and chosen to satisfy the Courant–Friedrichs–Lewy (CFL) condition, specifically \(\Delta t / h^2 < 1 / (6\pi)\), following the criterion proposed in Ref.~\cite{Chen2021}.

\subsection{Parameter Space}

To investigate the relaxation of the IG around a BECDM vortex line, we explore a three-dimensional parameter space \((g, M_{\mathrm{BEC}} / M_{\mathrm{IG}}, v_{\mathrm{m}})\), where \(M_{\mathrm{BEC}} / M_{\mathrm{IG}}\) denotes the mass ratio between the condensate and the gas.

We consider two values of the self-interaction strength: \(g = 1\) and \(g = 100\), to assess the impact of quantum pressure on vortex stability. The BECDM core mass is fixed at \(M_{\mathrm{BEC}} = 18.85\), while the mass ratio is varied over \(M_{\mathrm{BEC}} / M_{\mathrm{IG}} = 0.1\), 1, and 10. The maximum initial gas velocity is set to \(v_{\mathrm{m}} = 0.5\) and 1.0, controlling the dynamical response of the IG component.

The remaining parameters are held constant: the vortex number is set to \(m = 1\), corresponding to the most stable and energetically favorable configuration~\cite{Rindler_Daller_2012, Nikolaieva_2021}; the polytropic constant is \(K = 0.1\), setting a moderate internal energy scale; and the adiabatic index is \(\gamma = 5/3\), appropriate for a non-relativistic monoatomic gas. This setup yields a total of \(2 \times 3 \times 2 = 12\) simulations, designed to quantify how BECDM vortex lines influence the surrounding baryonic medium.

\subsection{Diagnostics}

To characterize the evolution of the system, we compute several macroscopic quantities for both the BECDM and IG components at each time step. For the BECDM we evaluate the following integrals:

\[
\begin{aligned}
    M_{\mathrm{BEC}} &= \int |\Psi|^2 \, d^3x, \\
    K_{\mathrm{BEC}} &= -\frac{1}{2} \int \Psi^* \nabla^2 \Psi \, d^3x, \\
    W_{\mathrm{BEC}} &= \frac{1}{2} \int |\Psi|^2 V \, d^3x, \\
    I_{\mathrm{BEC}} &= \frac{g}{2} \int |\Psi|^4 \, d^3x,
\end{aligned}
\]
where \(M_{\mathrm{BEC}}\) is the total mass of the condensate, \(K_{\mathrm{BEC}}\) its quantum kinetic energy, \(W_{\mathrm{BEC}}\) the gravitational potential energy, and \(I_{\mathrm{BEC}}\) the self-interaction energy associated with the nonlinear term.

For the IG component, we compute:
\[
\begin{aligned}
    M_{\mathrm{IG}} &= \int \rho \, d^3x, \\
    K_{\mathrm{IG}} &= \frac{1}{2} \int \rho |\vec{v}|^2 \, d^3x, \\
    W_{\mathrm{IG}} &= \frac{1}{2} \int \rho V \, d^3x, \\
    U_{\mathrm{IG}} &= \int \frac{p}{\gamma - 1} \, d^3x,
\end{aligned}
\]
where \(M_{\mathrm{IG}}\) is the total gas mass, \(K_{\mathrm{IG}}\) the bulk kinetic energy, \(W_{\mathrm{IG}}\) the gravitational potential energy, and \(U_{\mathrm{IG}}\) the internal energy.

From these quantities, we define the total energy and virial expressions for each component:
\[
\begin{aligned}
    E_{\mathrm{BEC}} &= K_{\mathrm{BEC}} + W_{\mathrm{BEC}}, \\
    Q_{\mathrm{BEC}} &= 2 K_{\mathrm{BEC}} + W_{\mathrm{BEC}} + 3 I_{\mathrm{BEC}}, \\
    E_{\mathrm{IG}}  &= K_{\mathrm{IG}} + W_{\mathrm{IG}} + U_{\mathrm{IG}}, \\
    Q_{\mathrm{IG}}  &= 2 K_{\mathrm{IG}} + W_{\mathrm{IG}} + 3 U_{\mathrm{IG}},
\end{aligned}
\]
where \(E_{\mathrm{BEC}}\) and \(E_{\mathrm{IG}}\) are the total energies of the condensate and gas respectively, whereas \(Q_{\mathrm{BEC}}\) and \(Q_{\mathrm{IG}}\) are virial scalars derived from the generalized virial theorem. In equilibrium, these indicators satisfy \(Q \approx 0\), providing a measure of dynamical relaxation and stability for each component.

\section{Results}
\label{sec:results}

\subsection{Vortex stability}

The stability of isolated vortex-line configurations in BECDM has been previously investigated in~\cite{Nikolaieva_2021}, their interaction with central black holes in~\cite{Glennon:2023oqa}, and their dynamics during head-on mergers in~\cite{sivakumar2025revealingturbulentdarkmatter}. In this work, we analyze vortex-line stability under a qualitatively different kind of perturbation: one induced by gravitational condensation of a randomly initialized ideal gas (IG). This scenario serves as a dynamical testbed to evaluate the robustness of vortex-line configurations when embedded in a collapsing baryonic medium.

\begin{figure}
    \centering
    \includegraphics[width=8cm]{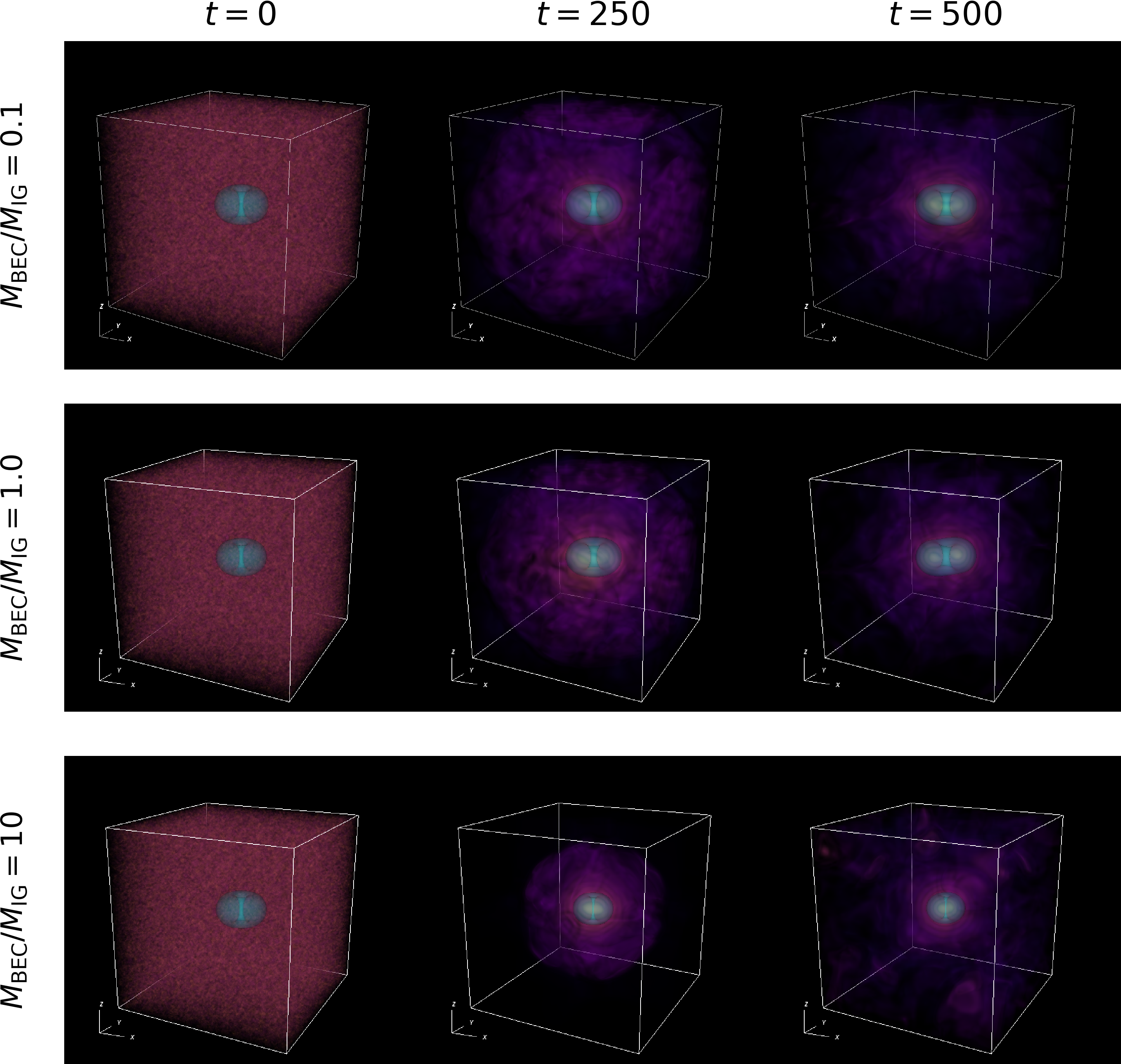}
    \caption{
    Evolution of a vortex-line configuration embedded in an ideal gas undergoing gravitational condensation. Each panel shows a 3D volume rendering of the BECDM density, with a cyan isosurface highlighting the vortex core, overlaid on a colormap of the IG specific internal energy. Columns correspond to simulation times \( t = 0 \), 250, and 500, and rows to mass ratios \( M_{\mathrm{BEC}} / M_{\mathrm{IG}} = 0.1 \), 1.0, and 10.0. The self-interaction strength is set to \( g = 1 \), and the maximum initial gas velocity is \( v_m = 0.5 \). The vortex-line remains stable in all cases, despite the inhomogeneous and time-dependent baryonic background.}
    \label{fig:vortex3D}
\end{figure}

Figure~\ref{fig:vortex3D} illustrates the evolution of the system at \(t = 0\), 250, and 500. Each panel shows a three-dimensional rendering of the BECDM density, with a central isosurface highlighting the vortex line, overlaid on a volumetric colormap of the IG's specific internal energy. From top to bottom, the rows correspond to mass ratios \(M_{\mathrm{BEC}} / M_{\mathrm{IG}} = 0.1\), 1.0, and 10.0, all for the case \(g = 1\) and maximum initial gas velocity \(v_m = 0.5\). The gas initially follows a homogeneous but randomly seeded distribution, and progressively condenses due to gravitational attraction.

Remarkably, despite the asymmetry of the baryonic collapse, the vortex-line structure in the BECDM remains well defined throughout the simulation. This supports the hypothesis that vortex lines are not merely stationary solutions but dynamically stable attractors under moderate baryonic perturbations.

\begin{figure}
    \centering
    \includegraphics[width=8cm]{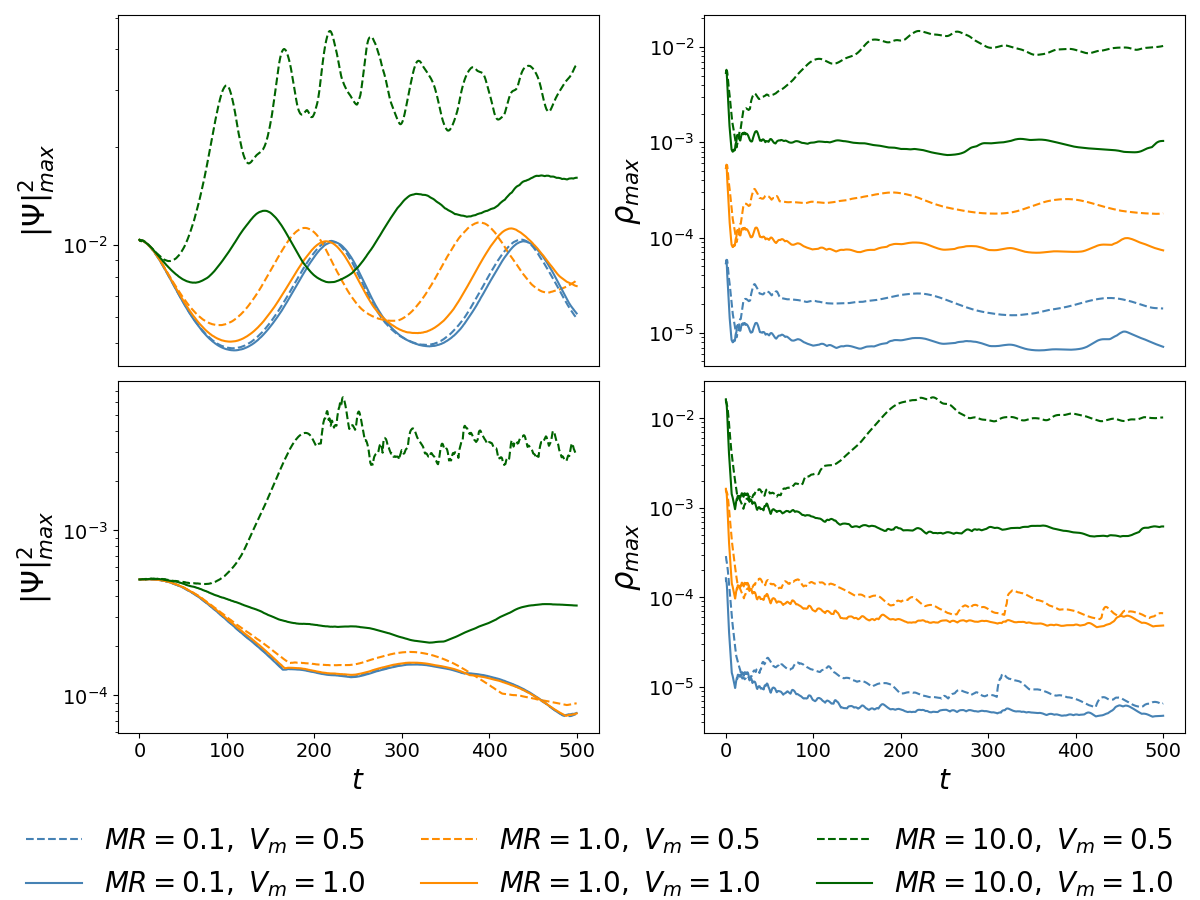}
    \caption{
    Time evolution of the maximum densities $|\Psi|^2$ and $\rho$ for the BECDM  and IG components, for two self-interaction strengths: \( g = 1 \) (top row) and \( g = 100 \) (bottom row). Solid lines correspond to \( v_m = 1.0 \), and dashed lines to \( v_m = 0.5 \). The three colors represent mass ratios \( M_{\mathrm{BEC}} / M_{\mathrm{IG}} = 0.1 \) (blue), \( M_{\mathrm{BEC}} / M_{\mathrm{IG}} = 1.0 \) (orange), and \( M_{\mathrm{BEC}} / M_{\mathrm{IG}} = 10.0 \) (green). For \( g = 1 \), all configurations display stable behavior, with bounded oscillations in the BECDM peak and smooth relaxation of the gas. In contrast, most cases with \( g = 100 \) show signs of instability, such as condensate decay and irregular fluctuations, except for the most massive and coldest configuration (\( M_{\mathrm{BEC}} / M_{\mathrm{IG}} = 10 \), \( v_m = 0.5 \)), which retains a stable core.}
    \label{fig:rhomax}
\end{figure}

A more quantitative assessment is presented in Figure~\ref{fig:rhomax}, which shows the evolution of the maximum densities of both components for two values of the self-interaction strength, \(g = 1\) and \(g = 100\). The left column corresponds to the BECDM component, and the right to the IG. Each curve represents a different mass ratio (\(M_{\mathrm{BEC}} / M_{\mathrm{IG}} = 0.1\), 1.0, and 10.0), with dashed and solid lines indicating \(v_m = 0.5\) and \(v_m = 1.0\), respectively.

For \(g = 1\), all configurations exhibit bounded oscillations in the BECDM maximum density, indicating that the vortex-line structure remains dynamically stable across the parameter space explored. The gas component settles into a quasi-stationary state after an initial transient, reaching higher equilibrium densities for larger mass ratios.

In contrast, the behavior for \(g = 100\) is markedly different. Most simulations with this value show strong density fluctuations, decay of the BECDM peak, and clear signs of instability—particularly for low or moderate mass ratios. The only exception is the case with \(M_{\mathrm{BEC}} / M_{\mathrm{IG}} = 10\) and \(v_m = 0.5\), which maintains a relatively stable and compact core throughout the simulation. These results indicate that strong self-interaction does not universally enhance vortex stability; in some regimes, it may even trigger dynamical instabilities in the presence of baryonic perturbations.

Overall, the persistence of vortex structures depends not only on the self-interaction strength, but also on the gravitational dominance of the BECDM component and the amount of kinetic energy injected by the gas.

\subsection{Ideal gas condensation}

The vortex-line configuration not only remains dynamically stable in the presence of a gravitationally coupled ideal gas, but also plays a central role in guiding its condensation. As shown in Figure~\ref{fig:vortex3D}, despite the initially random distribution of the gas, the baryonic component collapses preferentially around the preexisting BECDM vortex core. This behavior happens for all mass ratios $M_{\mathrm{BEC}} / M_{\mathrm{IG}}$, although the intensity of the collapse depends on the mass ratio and the initial velocity dispersion.

The gravitational potential induced by the BECDM vortex line acts as a condensation seed, drawing the gas toward its center, where it accumulates and thermalizes. As illustrated in Figure~\ref{fig:rhomax}, the gas density undergoes an initial phase of rapid compression, followed by partial stabilization into a quasi-stationary state in most cases.

For \( g = 1 \), the gas settles into a relatively steady central density after the initial collapse, with higher equilibrium values observed for larger mass ratios \( M_{\mathrm{BEC}} / M_{\mathrm{IG}} \), reflecting the stronger gravitational confinement exerted by a more massive condensate. In contrast, for \( g = 100 \), the gas exhibits more erratic dynamics: only the configuration with \( M_{\mathrm{BEC}} / M_{\mathrm{IG}} = 10 \) and \( v_m = 0.5 \) achieves a stable state, while the remaining cases show persistent oscillations and transient density peaks, driven by the instability of the underlying vortex.

Despite these fluctuations, the gas maximum density remains bounded in all simulations, indicating that the condensation process does not lead to runaway collapse. These results suggest that BECDM vortex lines can act as gravitational traps for baryonic matter, potentially facilitating the formation of visible structures. However, the efficiency and stability of this mechanism depend critically on the balance between condensate dominance and gas dynamics, and are not necessarily enhanced by stronger bosonic self-interactions.

\begin{figure}
    \centering
    \includegraphics[width=8cm]{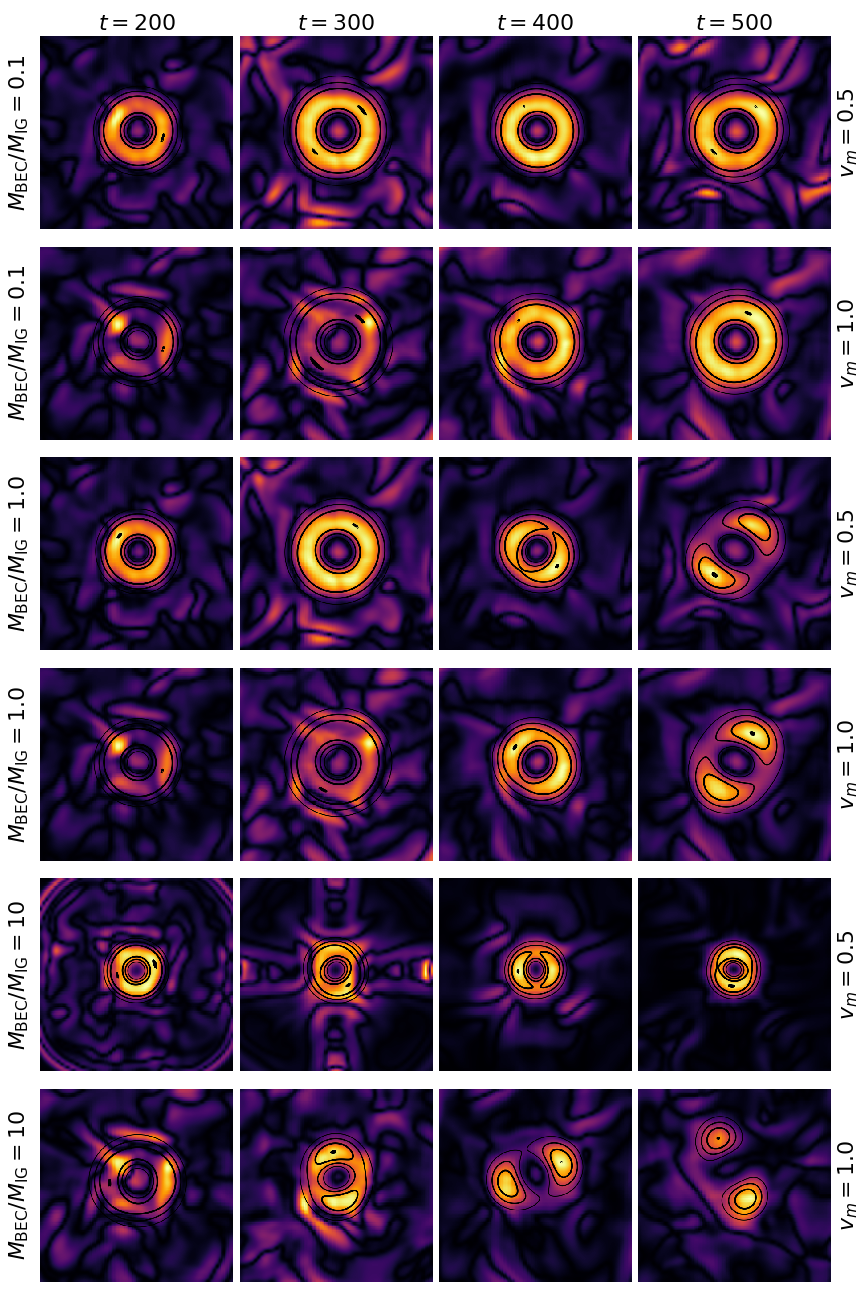}
    \caption{
    Laplacian of Gaussian (LoG)-filtered density maps of the ideal gas in the \(z=0\) plane at times \(t = 200\), 300, 400, and 500 (columns), for different mass ratios and initial gas velocities (rows). The background colormap represents the magnitude of the LoG-filtered gas density, highlighting local gradients and edge-like features. Overlaid black contours correspond to isodensity levels of the BECDM component for \(g = 1\). In most cases, particularly for \(M_{\mathrm{BEC}} / M_{\mathrm{IG}} \geq 1.0\), the filtered gas exhibits ring-like or distorted structures that align spatially with the vortex-core morphology of the condensate.}
    \label{fig:densitylog_g1}
\end{figure}

\begin{figure}
    \centering
    \includegraphics[width=8cm]{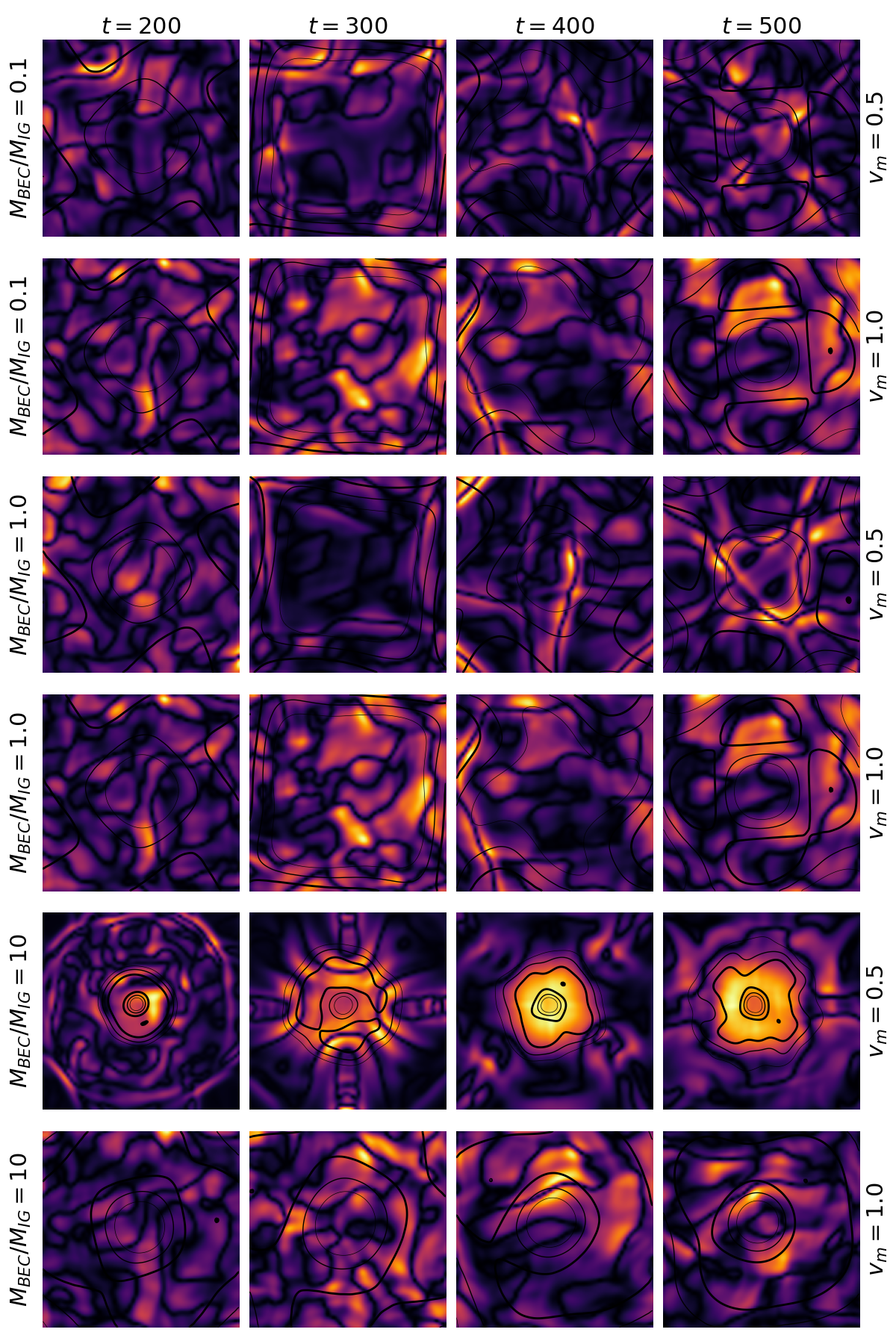}
    \caption{
    Similar to Figure~\ref{fig:densitylog_g1}, but for a self-interaction strength \(g = 100\). In this case, most configurations exhibit weaker morphological correspondence between the gas and BECDM components, due to increased internal pressure in the condensate, which inhibits coherent baryonic condensation. An exception occurs for \(M_{\mathrm{BEC}} / M_{\mathrm{IG}} = 10\) and \(v_m = 0.5\), where the gas collapses effectively and reinforces the central gravitational potential, allowing the condensate to remain confined and imprint a potentially detectable signature.}
    \label{fig:densitylog_g100}
\end{figure}

To assess whether the vortex line leaves a detectable morphological imprint on the gas, Figures~\ref{fig:densitylog_g1} and~\ref{fig:densitylog_g100} present 2D slices at \( z = 0 \) of the LoG-filtered gas density, with superimposed isocontours of the BECDM density. The colormap displays the filtered gas distribution, enhancing edge structures and localized gradients, while the black contours trace regions of constant BECDM density. For a detailed explanation of the Laplacian-of-Gaussian (LoG) operator and its implementation in this context, we refer the reader to Appendix~\ref{app:LoG}.

The results show that, despite the chaotic nature of the gas, the vortex-line structure consistently correlates with ring-like or crescent-shaped features in the filtered gas density. These correlations become more pronounced at late times (\( t \geq 400 \)) and for higher mass ratios. In particular, for \( M_{\mathrm{BEC}} / M_{\mathrm{IG}} = 1.0 \) and \( M_{\mathrm{BEC}} / M_{\mathrm{IG}} = 10.0 \), the LoG filter reveals symmetry-breaking patterns and ring distortions in the gas that spatially align with the BECDM vortex isocontours.

This suggests that, although the gas remains dynamically distinct and non-coherent at the quantum level, it responds gravitationally to the underlying condensate topology. As a result, the vortex line leaves a measurable imprint in the morphology of the baryonic component, potentially enabling indirect detection through luminous tracers.

\subsection{Diagnostics}
\label{sec:energetics}

\begin{figure}
    \centering
    \includegraphics[width=\linewidth]{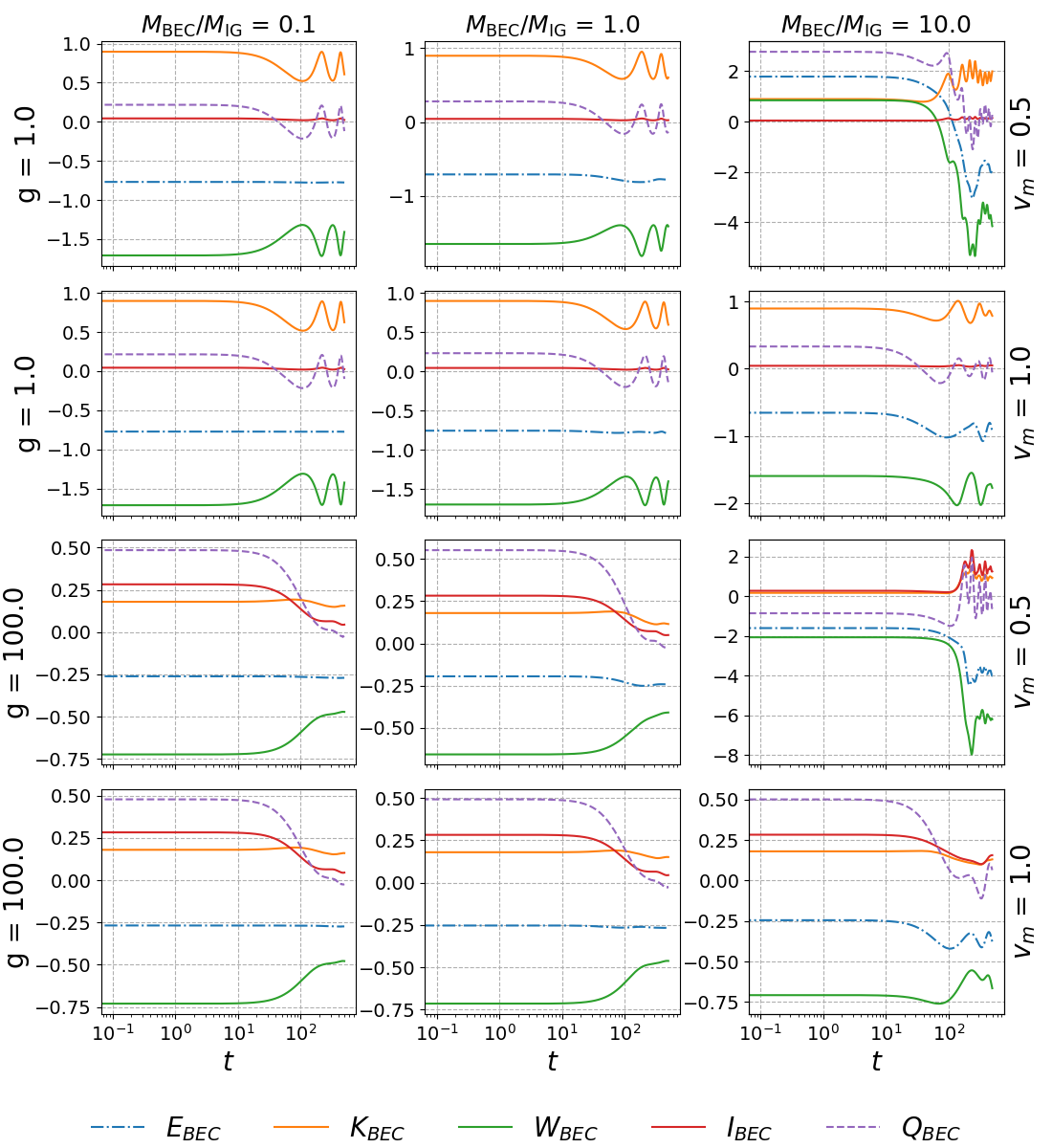}
    \caption{
    Evolution of the energy components and \( Q_{\mathrm{BEC}} \) for the BECDM component, across all combinations of \( g \), \( \mathrm{M_{\mathrm{BEC}} / M_{\mathrm{IG}}} \), and \( v_m \). Each row corresponds to a fixed \( g \), and each column to a fixed mass ratio. Solid and dashed lines distinguish between \( v_m = 1.0 \) and \( v_m = 0.5 \), respectively. For low \( g \), the condensate displays virialized oscillatory behavior; in contrast, most high-\( g \) cases exhibit energy drift and virial imbalance, except for the most massive and coldest configuration. A logarithmic time axis is used to capture the early-time transients in the energy evolution.}
    \label{fig:EnergiesBEC}
\end{figure}

\begin{figure}
    \centering
    \includegraphics[width=\linewidth]{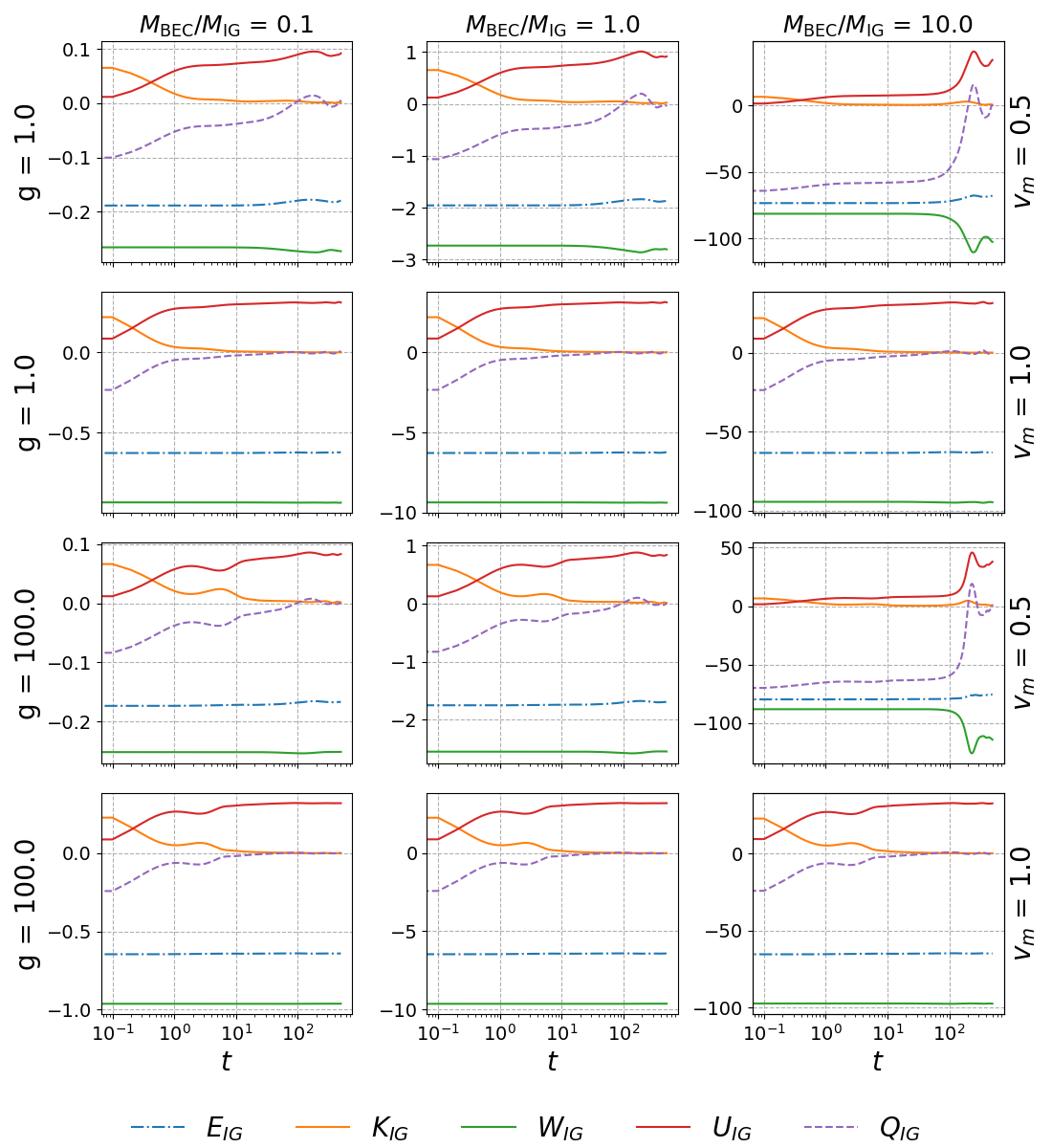}
    \caption{
    Evolution of the energy components and  \( Q_{\mathrm{IG}} \) for the IG component. Internal, kinetic, gravitational, and total energies are shown. Rows correspond to increasing self-interaction strength \( g \), and columns to increasing \( \mathrm{M_{\mathrm{BEC}} / M_{\mathrm{IG}}} \). The strongest gravitational coupling occurs for \( \mathrm{M_{\mathrm{BEC}} / M_{\mathrm{IG}}} = 10 \), where the gas virial combination departs significantly from equilibrium in the presence of vortex instability. A logarithmic time axis is used to capture the early-time transients in the energy evolution.}
    \label{fig:EnergiesIG}
\end{figure}

To further quantify the dynamical evolution and stability of the system, we analyze the time evolution of the energy components and virial quantities for both the BECDM and IG components. Figures~\ref{fig:EnergiesBEC} and~\ref{fig:EnergiesIG} present the results for all combinations of \( g \), \( \mathrm{M_{\mathrm{BEC}} / M_{\mathrm{IG}}} \), and \( v_m \), showing the kinetic (\( K \)), gravitational potential (\( W \)), internal (\( U \)), self-interaction (\( I \)), total energy (\( E \)), and virial scalar (\( Q \)) as defined in Section~\ref{sec:model}.

Figure~\ref{fig:EnergiesBEC} shows the evolution of the BECDM energies. For \( g = 1 \), the system exhibits oscillations around \( Q_{\mathrm{BEC}} \approx 0 \), particularly in high mass ratio scenarios, consistent with a dynamically stable vortex configuration. The total energy \( E_{\mathrm{BEC}} \) remains nearly constant over time, confirming global conservation. In contrast, for \( g = 100 \), most configurations display monotonic drifts in \( Q_{\mathrm{BEC}} \) and a reorganization of the energy components, signaling the onset of instabilities and a gradual loss of coherence—except for the case \( M_{\mathrm{BEC}} / M_{\mathrm{IG}} = 10 \), \( v_m = 0.5 \), which remains marginally stable.

The energies of the IG component, shown in Figure~\ref{fig:EnergiesIG}, reflect its gravitational response to the condensate. For \( g = 1 \), the gas reaches a quasi-stationary state where kinetic and internal energies stabilize, and the virial condition \( Q_{\mathrm{IG}} \approx 0 \) is approximately satisfied. However, for \( g = 100 \), most cases show clear virial imbalance and energy growth at late times especially for large \( M_{\mathrm{BEC}} / M_{\mathrm{IG}} \) which indicates a sustained dynamical activity triggered by the condensate's instability. Notably, configurations with \( M_{\mathrm{BEC}} / M_{\mathrm{IG}} = 10 \) exhibit significant increase in both internal and kinetic energy, confirming strong gravitational coupling between the components.

Overall, these diagnostics reinforce the conclusion that vortex-induced baryonic condensation is most efficient—and potentially observable—when the condensate is both massive and dynamically stable. Moreover, they show that strong bosonic self-interaction does not necessarily enhance stability, and may instead drive complex energy redistribution within the system.

\section{Conclusions}
\label{sec:conclusions}

In this work we investigate the dynamical stability and astrophysical role of vortex-line configurations in BECDM under gravitational interaction with an IG component. Our simulations explore a range of mass ratios, velocity dispersions, and bosonic self-interaction strengths, allowing us to assess the robustness of these structures in complex, collapsing environments.

We find that vortex lines in BECDM exhibit remarkable stability across a wide region of parameter space, particularly for moderate self-interaction strength (\( g = 1 \)). Even when embedded in a randomly seeded, collapsing baryonic medium, the vortex structure persists and remains well defined throughout the evolution. This highlights the role of vortex lines not merely as stationary solutions, but as dynamical attractors under realistic perturbations.

Beyond their stability, vortex lines act as gravitational seeds for baryonic condensation. Despite the initially random gas distribution, the baryonic component consistently collapses toward the vortex core, forming quasi-stationary structures that reflect the morphology of the underlying condensate. Notably, vortex-induced features in filtered gas density maps suggest a potential avenue for indirect detection through luminous tracers.

Our results also reveal a nontrivial dependence on the bosonic self-interaction strength. Contrary to intuitive expectations, increasing \( g \) does not universally enhance vortex stability. For \( g = 100 \), most configurations display clear signs of instability, including large density fluctuations and a breakdown of virial equilibrium—except in the most massive and coldest case studied. This demonstrates that strong self-interactions can destabilize the system in the presence of baryonic perturbations, challenging standard assumptions about structure formation in BECDM.

Energy diagnostics support this picture: stable configurations exhibit virialization and conservation of total energy, while unstable ones show clear energy redistribution. Importantly, the efficiency of baryonic condensation correlates with the dynamical stability of the condensate, indicating a tight coupling between dark and luminous matter dynamics in this framework.

Overall, our findings establish vortex lines in BECDM as dynamically robust structures capable of influencing baryonic evolution and potentially imprinting observable signatures. These results position vortex structures not merely as theoretical solutions, but as relevant, testable features in the nonlinear dynamics and observational phenomenology of fuzzy dark matter.

\section*{Acknowledgments}

Iv\'an \'Alvarez  and Carlos Tena receives support from the CONAHCyT graduate scholarship program. This research is supported by grants CIC-UMSNH-4.9, Laboratorio Nacional de C\'omputo de Alto Desempe\~no Grant Nos. 1-2024 and 5-2025.

\bibliography{BECDM}


\appendix

\section{Vortex Lines: Ground-State Solution}
\label{app:ground_state}

We consider the dimensionless, stationary GPP subsystem Eqs.~\eqref{eq:Schrodinger_adim} and \eqref{eq:Poisson_adim} in cylindrical coordinates \((r_{\perp}, \varphi, z)\), where the macroscopic wave function is assumed to take the form

\begin{equation}
    \psi(r_{\perp}, \varphi, z) = \phi(r_{\perp}, z) \, e^{i m \varphi},
\end{equation}

\noindent with \(m \in \mathbb{Z}\) denoting the quantized angular momentum or vortex number. This ansatz ensures single-valuedness and periodicity in the azimuthal direction. Substituting into the Gross-Pitaevskii equation yields a nonlinear eigenvalue problem:
\begin{equation}
    H \phi = \omega \phi,
    \label{eq:stationaryGP}
\end{equation}

\noindent where \(\omega\) is the eigenvalue and \(H\) the effective Hamiltonian, defined as:

\begin{equation}
    H = -\frac{1}{2} \left( \frac{\partial^2}{\partial r_{\perp}^2} + \frac{1}{r_{\perp}} \frac{\partial}{\partial r_{\perp}} + \frac{\partial^2}{\partial z^2} - \frac{m^2}{r_{\perp}^2} \right) + V + g \phi^2.
\end{equation}

\noindent The gravitational potential \(V(r_{\perp}, z)\) is determined self-consistently from Poisson equation:

\begin{equation}
    \frac{\partial^2 V}{\partial r_{\perp}^2} + \frac{1}{r_{\perp}} \frac{\partial V}{\partial r_{\perp}} + \frac{\partial^2 V}{\partial z^2} = \phi^2.
    \label{eq:Poisson_StationaryGP}
\end{equation}

\noindent We impose the following boundary conditions: \(\phi(0,z) = 0\) to enforce the vortex core along the \(z\)-axis; \(\partial_z \phi(r_{\perp},0) = 0\) to ensure equatorial symmetry; and asymptotic decay at infinity:
\[
\lim_{\sqrt{r_{\perp}^2 + z^2} \to \infty} \phi = \lim_{\sqrt{r_{\perp}^2 + z^2} \to \infty} \partial_{r_{\perp}} \phi = \lim_{\sqrt{r_{\perp}^2 + z^2} \to \infty} \partial_z \phi = 0.
\]
These conditions guarantee normalizability and, in the linear case (\(g = 0\)), uniqueness of the solution. For the gravitational potential, regularity and symmetry are imposed via \(\partial_{r_{\perp}} V(0,z) = 0\) and \(\partial_z V(r_{\perp},0) = 0\), respectively.

\smallskip

The system is solved numerically using the \textit{Imaginary Time Evolution Method} (ITEM), where the time-dependent GPP equation is evolved under the substitution \(\hat{t} = -i t\), yielding a diffusion-like equation:
\begin{equation}
    \frac{\partial \phi}{\partial \hat{t}} = H \phi.
    \label{eq:ITEM}
\end{equation}

\noindent This evolution suppresses excited states and allows convergence to the ground state of the Hamiltonian. The same approach applies to the nonlinear case (\(g \neq 0\)), typically converging to a ground or metastable configuration.

At each time step, the wave function is normalized to preserve the total mass,
\begin{equation}
    M = 2\pi \int \phi^2 \, r_{\perp} \, dr_{\perp} \, dz,
\end{equation}
ensuring convergence to a unique eigenvalue \(\omega\) for the chosen mass. Simultaneously, the potential \(V\) is updated by solving Eq.~\eqref{eq:Poisson_StationaryGP} via the \textit{Successive Over-Relaxation} (SOR) method to accelerate convergence.

The domain is finite: \((r,z) \in [0, R]^2\), with Neumann boundary conditions \(\partial_r \phi(R, z) = 0\) and \(\partial_z \phi(r, R) = 0\), which approximate asymptotic decay. To validate this approximation, we explore various values of \(R\) and assess boundary sensitivity for the case \(M = 2\pi\), \(g = 0\).

\begin{figure}
    \centering
    \includegraphics[width=8cm]{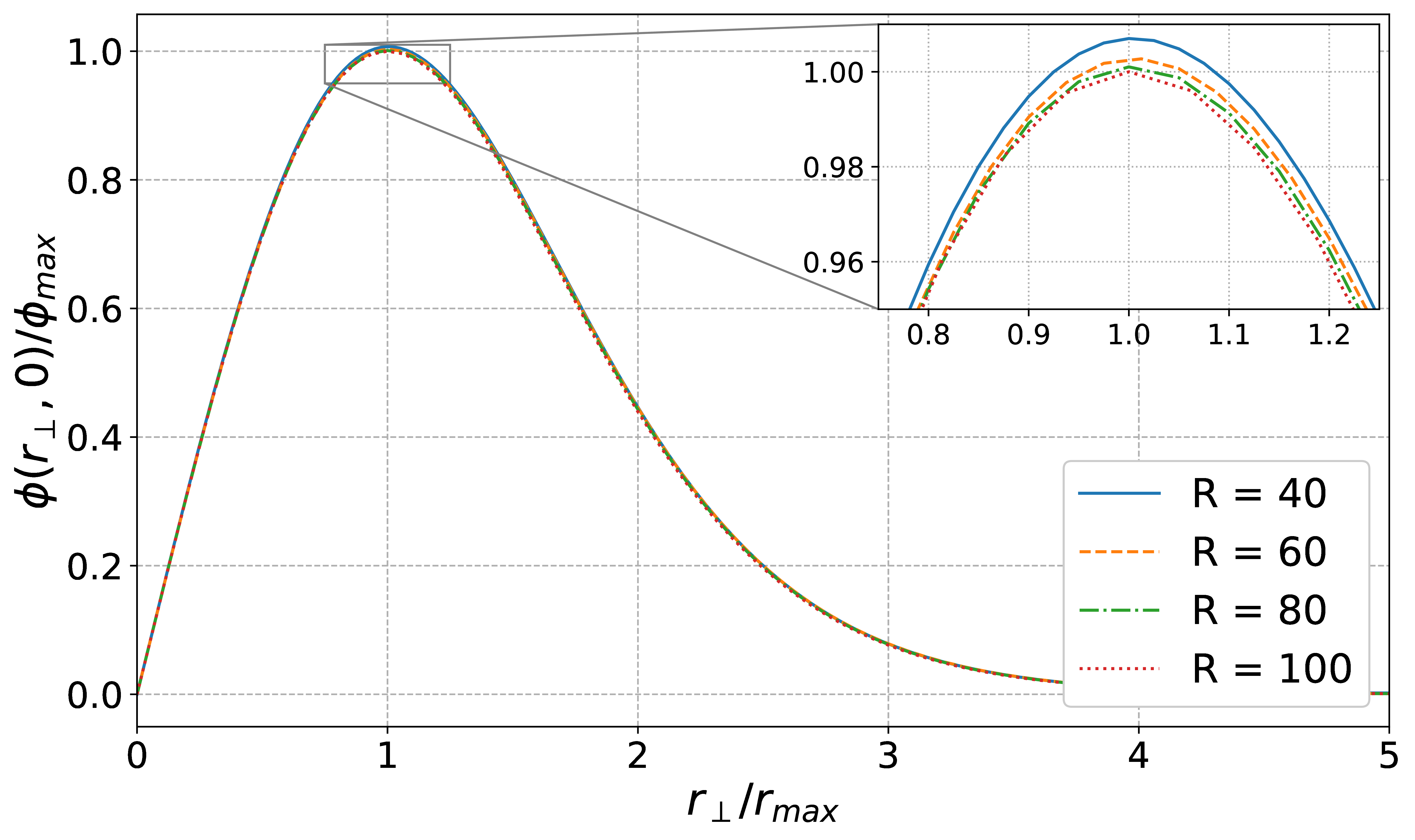}
    \caption{
    Radial profiles of the wave function \(\phi(r_{\perp}, z = 0)\) for different computational domain sizes \(R = 40\), 60, 80, and 100, with vortex number \(m = 1\), total mass \(M = 2\pi\), and \(g = 0\). The convergence of the profiles with increasing \(R\) confirms that Neumann boundary conditions accurately approximate the asymptotic behavior. All curves are normalized by \(\phi_{\max} \approx 0.0422\) at \(r_{\max} \approx 8.000\), making the solutions invariant under rescalings of the length scale \(\lambda\).}
    \label{fig:PhiZoom}
\end{figure}

Figure~\ref{fig:PhiZoom} displays radial profiles \(\phi(r_{\perp}, z=0)\) for \(R = 40\), 60, 80, and 100, with \(m = 1\). The profiles converge with increasing domain size, indicating that \(R = 40\) already provides a good approximation of the infinite-domain solution.

All solutions are normalized by the maximum amplitude \(\phi_{\max} \approx 0.0422\), occurring at \(r_{\max} \approx 8.000\), for \(M = 2\pi\) and \(g = 0\). This normalization renders the profiles invariant under changes in the length scale \(\lambda\), as discussed in~\cite{GuzmanUrena2004, Mocz:2017wlg, AlvarezGuzman2022}, ensuring uniqueness within the adopted scaling.

\subsection{Impact of the vortex number \texorpdfstring{$m$}{m}}

The vortex number \(m\) determines the quantized angular momentum of the condensate and has a direct impact on the core structure of the vortex. As \(m\) increases, the centrifugal barrier term \(m^2/r_{\perp}^2\) strengthens, widening the central depletion in the density profile and giving rise to toroidal configurations with reduced central density. Consequently, the peak of the wave function shifts outward, enlarging the radial extent of the structure.

\begin{figure}
    \centering
    \includegraphics[width=8cm]{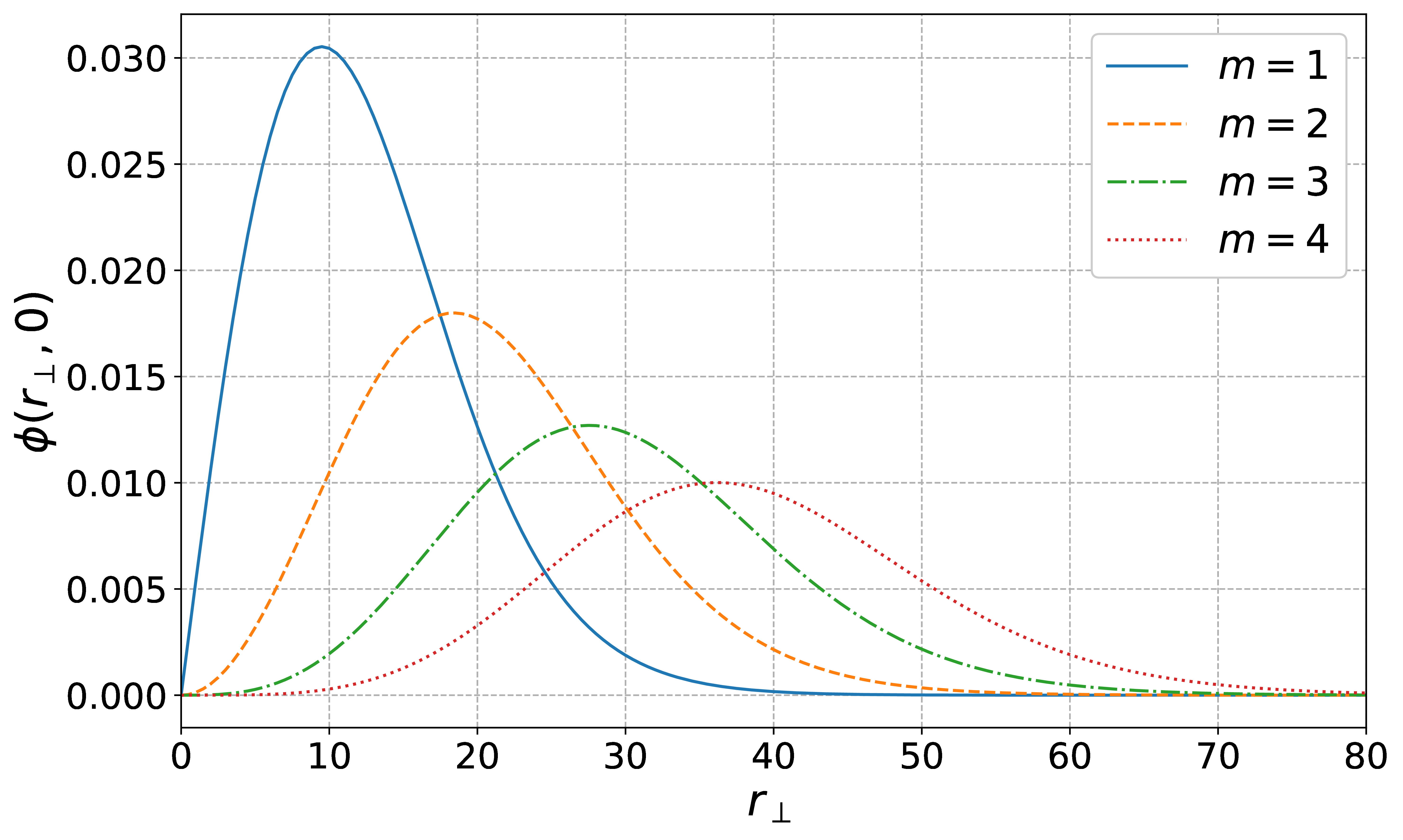}
    \caption{Radial profiles of the wave function \(\phi(r_{\perp}, z = 0)\) for different vortex numbers \(m\), with fixed self-interaction strength \(g = 10\) and total mass \(M = 2\pi\). Larger values of \(m\) lead to broader cores and outward-shifted maxima, due to the increased centrifugal barrier.}
    \label{fig:phi_vs_m}
\end{figure}

These effects are illustrated in Figure~\ref{fig:phi_vs_m}, which displays radial profiles of \(\phi(r_{\perp}, z = 0)\) for several values of \(m\), with fixed parameters \(g = 10\) and \(M = 2\pi\). As the vortex number increases, the profiles become progressively broader and flatter near the center, clearly revealing the influence of angular momentum.

\subsection{Impact of the self-interaction strength \texorpdfstring{$g$}{g}}
\label{Apend: Impact to g}

The parameter \(g\) governs the strength and nature of self-interactions within the condensate. For \(g > 0\), the interaction is repulsive, which favors more extended and less concentrated profiles. As \(g\) increases, the wave function becomes flatter and more diffuse, approaching the Thomas–Fermi regime. In contrast, for \(g < 0\), the interaction is attractive, promoting compact and high-density configurations, which may become unstable beyond a critical threshold of \(|g|\).

\begin{figure}
    \centering
    \includegraphics[width=8cm]{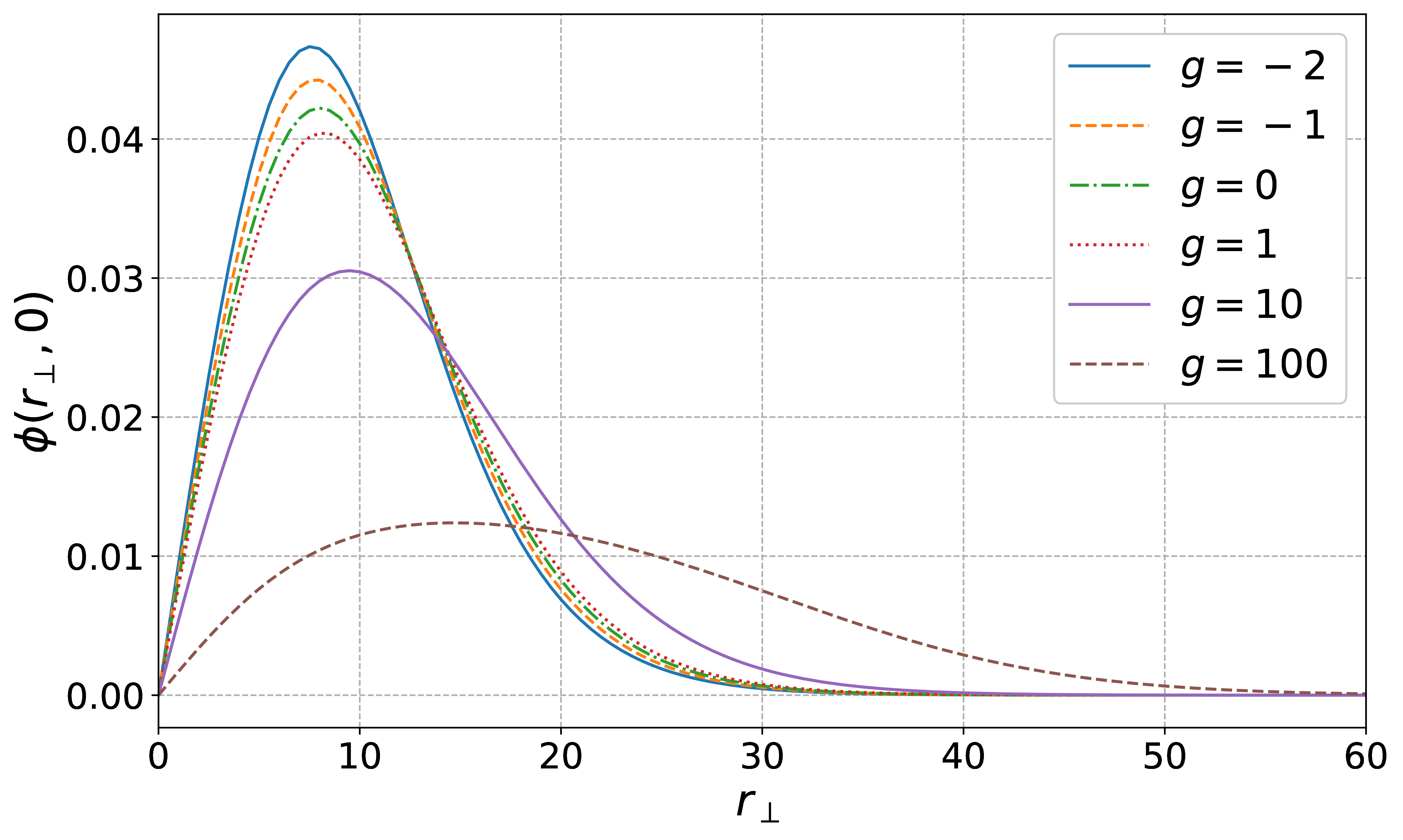}
    \caption{Radial profiles of the normalized wave function \(\phi(r_{\perp}, z=0)\) for different values of the self-interaction strength \(g\), with \(M = 2\pi\) and \(m = 1\). Increasing \(g\) results in broader, more diffuse profiles with reduced central amplitudes.}
    \label{fig:phi_vs_g}
\end{figure}

Figure~\ref{fig:phi_vs_g} presents radial profiles of the wave function for various values of \(g\), keeping \(M = 2\pi\) and \(m = 1\) fixed. As \(g\) increases, the profiles broaden and their central amplitudes decrease, illustrating the dispersion induced by repulsive self-interactions. On the contrary, for negative $g$ the configurations become more compact.

\subsection{Impact of the total mass \texorpdfstring{$M$}{M}}

The total mass \(M\) sets the overall normalization of the wave function and controls the gravitational binding of the configuration. Higher masses increase the self-gravity of the system, yielding more compact profiles with steeper central peaks. This effect competes with quantum pressure and self-interaction, shaping the equilibrium profile.

\begin{figure}
    \centering
    \includegraphics[width=8cm]{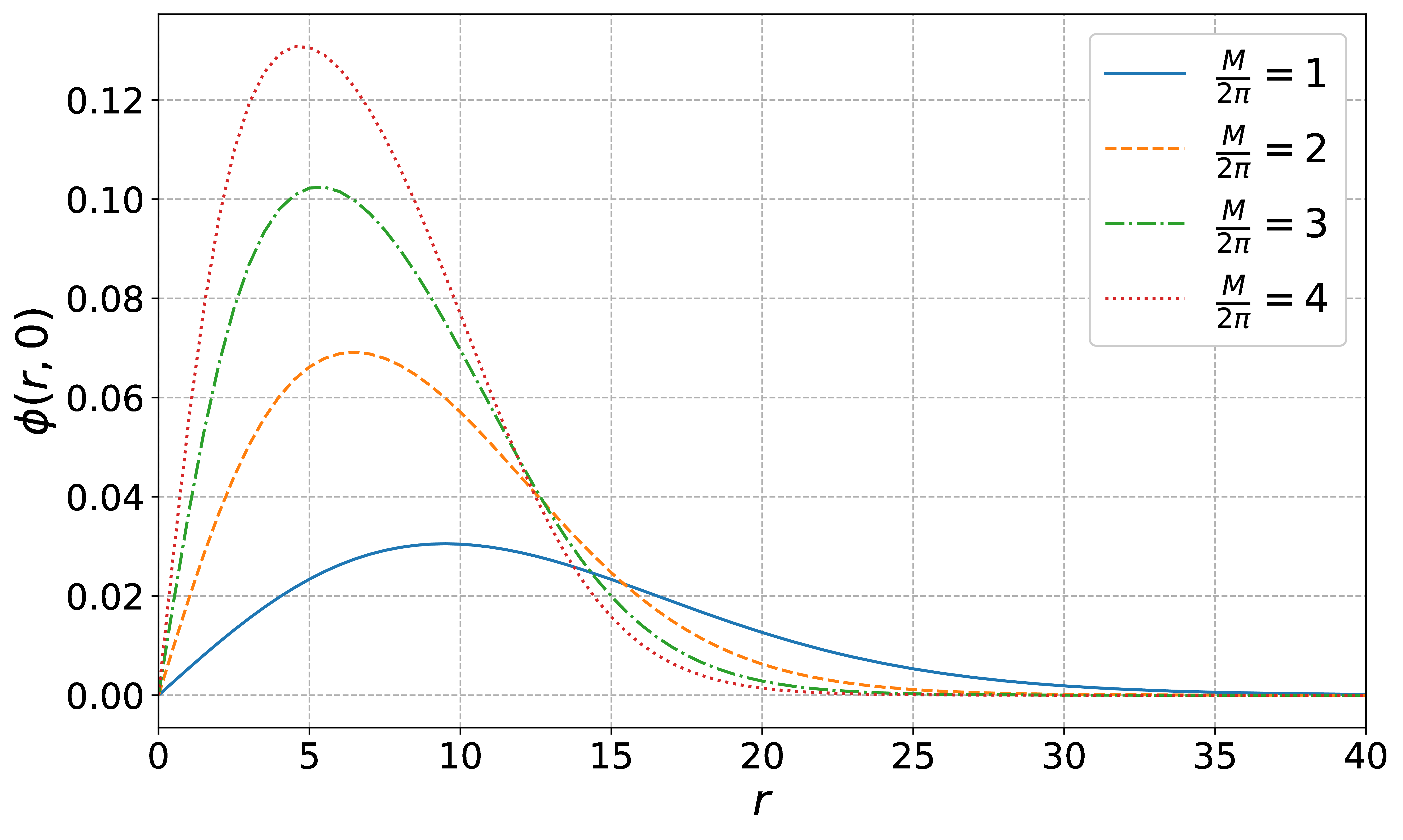}
    \caption{Radial profiles of the wave function \(\phi(r_{\perp}, z = 0)\) for different total masses \(M\), with \(m = 1\) and \(g = 10\). Increasing \(M\) enhances gravitational confinement, resulting in more compact and centrally peaked configurations.}
    \label{fig:phi_vs_M}
\end{figure}

Figure~\ref{fig:phi_vs_M} displays radial profiles for different total masses \(M\), with fixed parameters \(m = 1\) and \(g = 10\). As \(M\) increases, the core becomes more peaked and the profile more localized, indicating stronger gravitational confinement.

\section{Structure Detection via Laplacian-of-Gaussian Filtering}
\label{app:LoG}

To identify vortex-induced features in the projected gas density, we apply an edge-enhancement technique based on the Marr–Hildreth operator~\cite{MarrHildreth1980}, commonly known as the Laplacian-of-Gaussian (LoG) filter. This method is applied to the ideal gas density evaluated on the equatorial plane \( z = 0 \), and is defined as
\begin{equation}
    \mathrm{LoG}[\rho|_{z=0}] = \nabla^2_{(x,y)} \left( G_\sigma * \rho|_{z=0} \right) = \left( \nabla^2_{(x,y)} G_\sigma \right) * \rho|_{z=0},
\end{equation}
where \( G_\sigma(x, y) = \frac{1}{2\pi \sigma^2} \exp\left( -\frac{x^2 + y^2}{2\sigma^2} \right) \) is a Gaussian kernel of width \( \sigma \), the symbol \(*\) denotes convolution, and \( \nabla^2_{(x,y)} \) is the two-dimensional Laplacian in the \(x\text{-}y\) plane.

The LoG filter acts as a band-pass operator, it suppresses long-wavelength gradients and high-frequency noise, while enhancing intermediate-scale features. In our simulations, it effectively highlights ring-like depressions in the gas density associated with vortex-lines in the BECDM component.

We adopt a fixed filter width of \( \sigma = 1 \), which balances spatial resolution and noise suppression near the vortex core. To compute the filter efficiently, we work in Fourier space using the convolution theorem:

\begin{equation}
    \mathrm{LoG}[\rho|_{z=0}] = \mathcal{F}^{-1} \left[ \mathcal{F}(\nabla^2 G_\sigma) \cdot \mathcal{F}(\rho|_{z=0}) \right],
\end{equation}
where \( \mathcal{F} \) and \( \mathcal{F}^{-1} \) denote the two-dimensional Fourier and inverse Fourier transforms.

\begin{figure}
    \centering
    \includegraphics[width=0.9\linewidth]{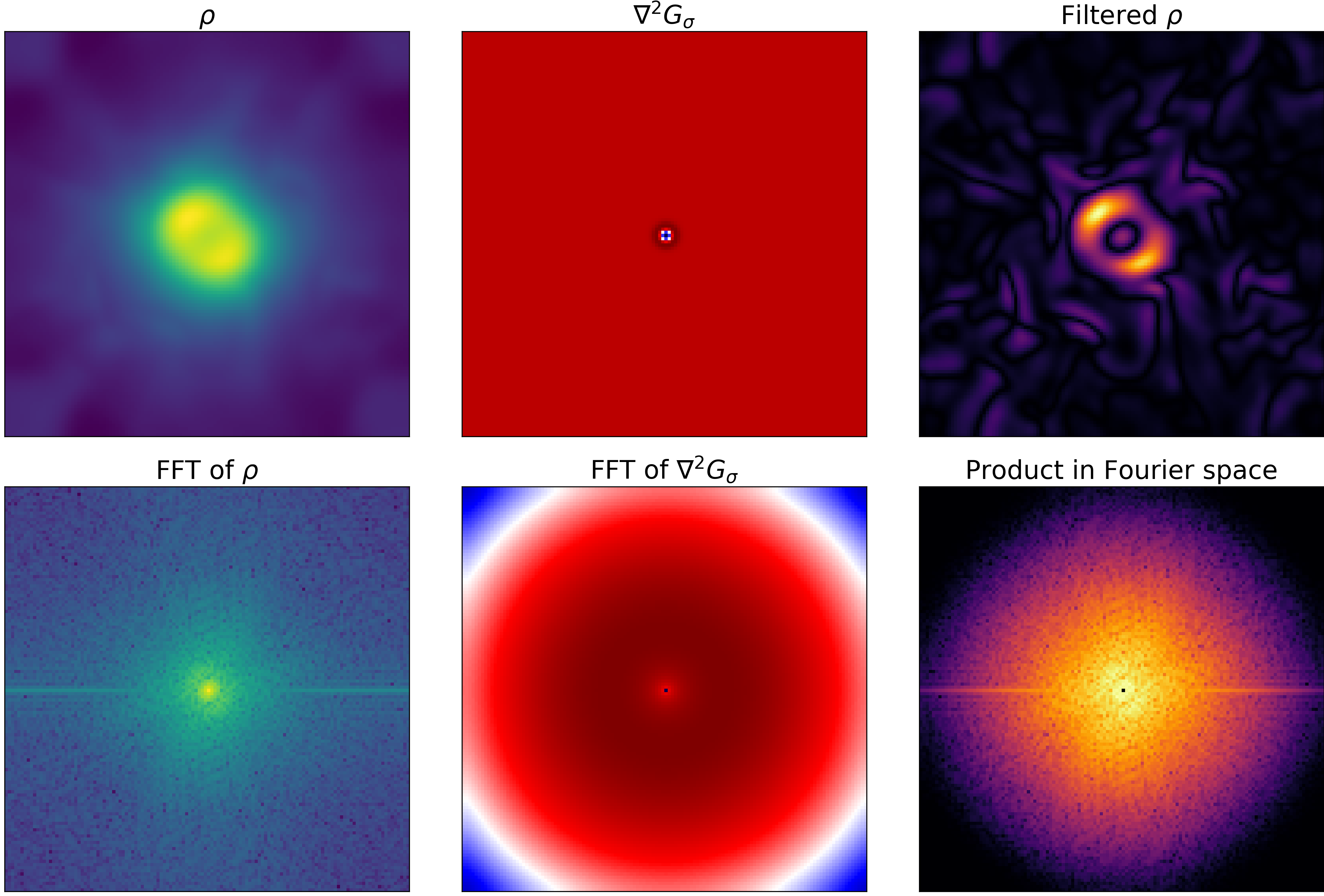}
    \caption{Application of the Laplacian-of-Gaussian (LoG) filter to the projected ideal gas density at \( z = 0 \). Top row: original density \( \rho \), LoG kernel \( \nabla^2 G_\sigma \), and the resulting filtered field. Bottom row: Fourier amplitudes for \( \rho \), \( \nabla^2 G_\sigma \), and their product. The filter uses \( \sigma = 1 \). Parameters: \( g = 1 \), \( v_m = 0.5 \), \( \mathrm{M_{\mathrm{BEC}} / M_{\mathrm{IG}}} = 1.0 \), \( t = 500 \). The ring seen in the filtered result coincides with the vortex core, confirming its physical origin.}
    \label{fig:LoG_example1}
\end{figure}

Figure~\ref{fig:LoG_example1} illustrates this procedure for a representative case with \( g = 1 \), \( v_m = 0.5 \), and \( \mathrm{M_{\mathrm{BEC}} / M_{\mathrm{IG}}} = 1.0 \) at \( t = 500 \), corresponding to the simulations in Section~\ref{sec:results}. The top row shows the unfiltered gas density, the LoG kernel, and the filtered result. The bottom row displays the corresponding Fourier amplitudes and their spectral product. The filtered image reveals a clear ring-like structure aligned with the BECDM vortex location, which is not easily discernible in the raw data.

To confirm that the detected ring is not a filtering artifact, we apply the same LoG procedure to a spherically symmetric collapse without vortex formation. This case corresponds to a homogeneous initial distribution of gas and BECDM. As discussed in~\cite{alvarezrios2024fermionbosonstarsattractorsfuzzy}, the system evolves into a Newtonian Fermion-Boson star with no angular momentum.

\begin{figure}
    \centering
    \includegraphics[width=0.9\linewidth]{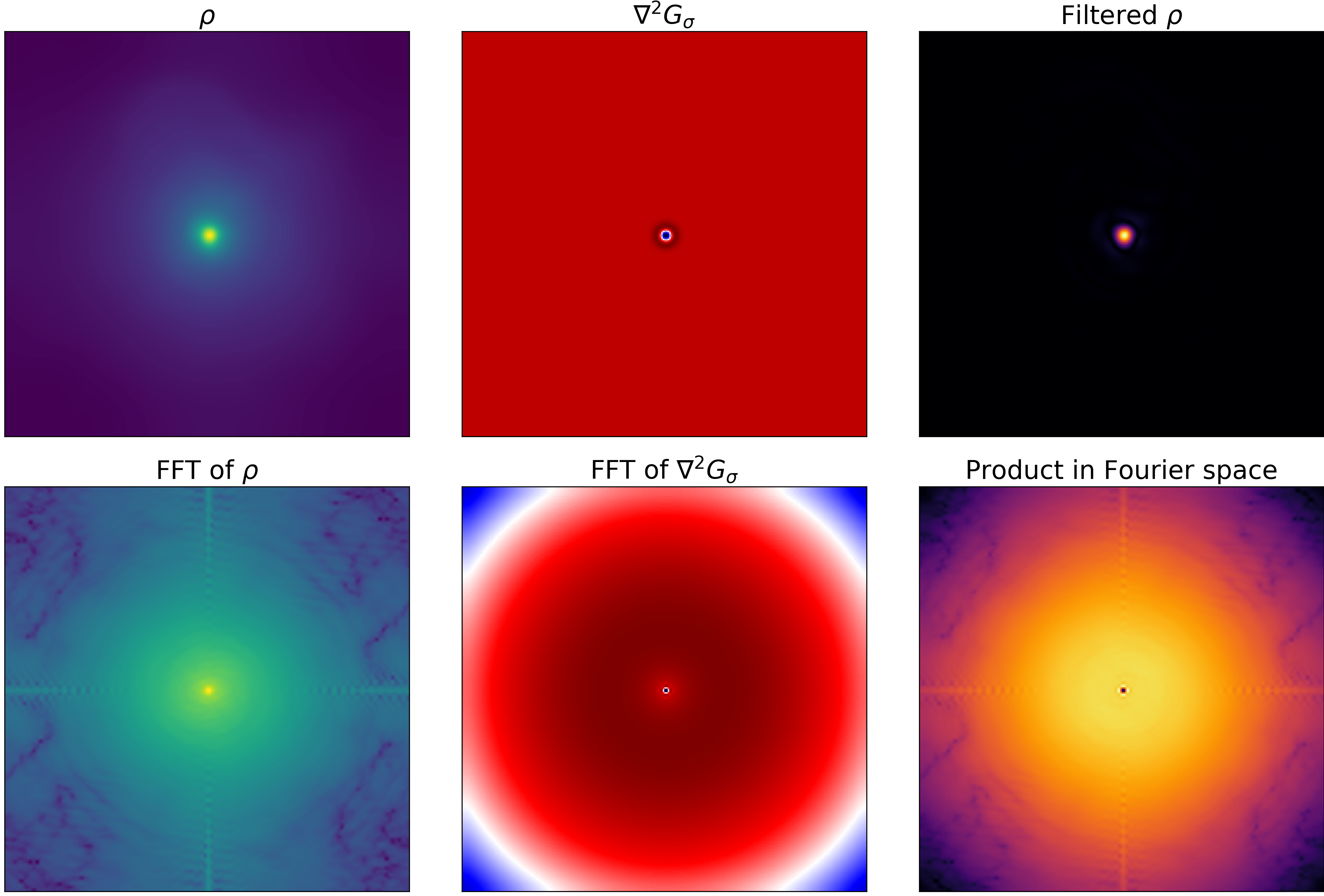}
    \caption{Laplacian-of-Gaussian filter applied to a spherically symmetric configuration without vortex formation. Top row: projected gas density, LoG kernel, and filtered result. Bottom row: Fourier transforms and spectral product. The absence of asymmetric features confirms that the vortex signature in Figure~\ref{fig:LoG_example1} is not a filtering artifact.}
    \label{fig:LoG_example2}
\end{figure}

Figure~\ref{fig:LoG_example2} shows that the filtered result lacks any ring-like or asymmetric structure. The LoG output remains spherically symmetric, consistent with a centrally peaked curvature response. This contrast confirms that the vortex-induced signal in Figure~\ref{fig:LoG_example1} is genuine and not an artifact of the filtering method.

\end{document}